\documentclass[floatfix,twocolumn,pre,showpacs]{revtex4}
\usepackage{graphicx}
\usepackage{subfigure}
\usepackage{amssymb,amsmath}

\begin{document}

\title{Swimming with a friend at low Reynolds number}

\author {C. M. Pooley, G. P. Alexander, and J. M. Yeomans*}

\affiliation{Rudolf Peierls Centre for Theoretical Physics, 1 Keble Road, Oxford, OX1 3NP, United Kingdom.}

\date{\today}


\begin{abstract}
We investigate the hydrodynamic interactions between microorganisms swimming at low Reynolds number. By considering simple model swimmers, and combining analytic and numerical approaches, we investigate the time-averaged flow field around a swimmer. At short distances the swimmer behaves like a pump. At large distances the velocity field depends on whether the swimming stroke is invariant under a combined time-reversal and parity transformation. We then consider two swimmers and find that the interaction between them consists of two parts; a {\em dead} term, independent of the motion of the second swimmer, which takes the expected dipolar form and a {\em live} term resulting from the simultaneous swimming action of both swimmers which does not. We argue that, in general, the latter dominates. The swimmer--swimmer interaction is a complicated function of their relative displacement, orientation and phase, leading to motion that can be attractive, repulsive or oscillatory. 
\end{abstract}
\maketitle

Because of their size, the swimming motion of microscopic and
mesoscopic organisms, such as bacteria and cells, corresponds to low
Reynolds number. As pointed out by Purcell~\cite{purcell77}, in this
regime swimming mechanisms are very different from those operative at
human length scales. In particular any swimming strategy must involve
a cyclic and non time reversible motion. Examples include the helical
motion of sperm flagellae~\cite{lighthill76}, the synchronised beating
of cilia in bacteria~\cite{guirao07}, and the Dreyfus
microswimmer~\cite{dreyfus05} where a colloid is driven by a tail of
magnetic particles actuated by an oscillating field.

Interesting progress has been made in understanding the propulsion of
individual swimmers~\cite{lighthill76,higdon79b,shapere89a,stone96}
and in developing optimal swimming
strategies~\cite{shapere89b,becker03,tam07}. However, much less is
understood about the collective behaviour of
microswimmers. There is considerable evidence that groups of swimming
cells undergo large density fluctuations and form transient 
patterns~\cite{toner05,goldstein04, guell88, budrene91, budrene95}. Indeed such collective
behaviour appears to be a generic feature of so-called active systems,
which produce their own energy~\cite{ramaswamy06}.

An important step towards understanding the behaviour of
many swimmers  is to calculate the hydrodynamic interaction between
two swimmers. Previous studies have investigated this  problem for a
number of special cases. For example, Ishikawa and
Hota~\cite{ishikawa06} performed experiments to find the
interactions between two swimming {\it Paramecium} cells. Computational
modelling of the hydrodynamic interaction between two flagella was
reported in~\cite{kim04,nasseri97}. Ishikawa {\it et
al.}~\cite{pedley06} modelled swimmers as spheres with prescribed
tangential velocities (known as squirmers). They calculated the far
field interaction between two squirmers and matched the results onto a
near field lubrication theory. 
An important disadvantage of this method is that it neglects any interaction 
resulting from the simultaneous motion of both swimmers.
We find that these neglected terms are, in fact, very important for the dynamical 
behaviour of a wide class of swimmers.
 

The aim of this article is to calculate two swimmer interactions, 
emphasising the importance of the details of the swimming
stroke. We present both analytic and numerical results based on a
minimal swimming model, first introduced by Najafi and
Golestanian~\cite{najafi04}. 
We first consider the average flow field around a swimmer. At
short distances the swimmer behaves like a pump which moves forward by
pushing fluid from in front to behind itself. At large distances
the average flow field around a swimmer is naively expected to be
dipolar ($\sim r^{-2}$). However, we argue that for a wide class of swimmers, which are
invariant under a combined time-reversal and parity (TP) transformation, the
leading order term vanishes and the far flow field has a $r^{-3}$
dependence. Examples of these swimmers include Purcell's
three link swimmer~\cite{purcell77}, the three sphere swimmer of Najafi and
Golestanian~\cite{najafi04}, the snake swimmer and a helical filament. Indeed the
$r^{-3}$ term in the velocity field will remain important for many
swimmers which are close to TP invariant.

We then consider two swimmers, Alice and Bob, say. We find that the
effect on Alice of Bob's swimming consists of two principal
components; a {\em  dead} term, which depends only on the flow field
generated by Bob and which takes the expected dipolar form, and a
{\em live} term resulting from the simultaneous swimming motion of both
Alice and Bob which does not. Interestingly, it is the latter 
which dominates, meaning that interactions become not only a
complicated function of relative displacement and 
orientation, but also of relative phase. We show that trajectories
of the two swimmers can be attractive, repulsive or
oscillatory.

Locomotion is initiated by shape changes in the swimmer, which induce forces acting on the surrounding fluid and 
thus create a flow. The fundamental solution of the Stokes' equations describing the response to a point force, 
${\bf f}$, is called a stokeslet and creates a fluid flow
\begin{equation}
u_i({\bf x}) = \tfrac{f_j}{8\pi \mu} \Bigl\{ \tfrac{\delta_{ij}}{r} + \tfrac{({\bf x}-{\bf y})_i({\bf x}-{\bf 
y})_j}{r^3} \Bigr\} \, ,
\label{eq:stokeslet}
\end{equation}
where ${\bf y}$ is the point of application of the force, $r=|{\bf x}-{\bf y}|$ and $\mu$ is the fluid viscosity. 
Since the Stokes' equations are linear, the fluid motion associated
with swimming may be given by a superposition
of such stokeslet flows, which at large distances may be expanded in successively higher powers of $r^{-1}$ and 
the far field flow approximated by the lowest order, non-zero term. Since the total force on the swimmer 
has to be zero the leading order term vanishes~\cite{lighthill76}. The next term, which varies as $r^{-2}$, is 
called a force dipole and is generally assumed to dominate the far field 
behaviour~\cite{ortiz05,simha02,pedley92}.

\begin{figure}
\begin{center}
\includegraphics[width=80mm]{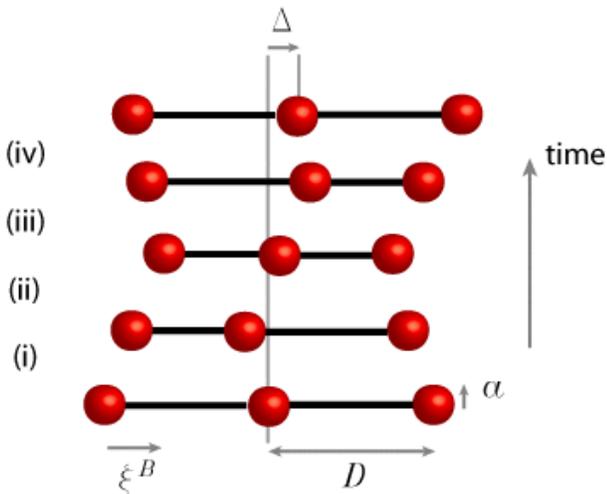}
\end{center}
\caption{The cyclic motion of the linear three sphere swimmer~\cite{najafi04}. The swimmer goes through the four 
steps (i) - (iv) and returns to its original shape, but displaced by a
distance $\Delta$ to the right. }
\label{fig:golepic}
\end{figure}

Indeed this is always true of the {\it instantaneous} fluid velocity.
However, an intriguing feature of the {\it average} flow is that the
force dipole term can be absent.
To show this we employ a simple model swimmer which
consists of three spheres of radius $a$ joined by thin rods, or
`arms'. It moves by shortening and extending these arms in a
periodic and time irreversible manner, as depicted in
Fig.~\ref{fig:golepic}. A unit vector ${\bf
n}$ pointing in the direction of motion, left to right in
Fig.~\ref{fig:golepic}, defines an orientation for the swimmer. 
The variable $D$ gives the extended arm length
and $\xi^F$ and $\xi^B$ denote the amplitudes of the strokes performed 
by the front and back arms respectively.  
Najafi and Golestanian~\cite{najafi04} obtained an analytic expression
for the velocity of the swimmer for $\xi^F = \xi^B$ using the Oseen tensor formulation of 
hydrodynamics, valid in the limit $a/D \ll 1$.

Using a similar approach, we have calculated the average flow field generated by the swimmer: 
\begin{multline}
\langle {\bf u} \rangle = \tfrac{a}{T} \Bigl\{ \bigl[ \tfrac{29 a\xi^B\xi^F(\xi^B-\xi^F)}{64 D^2} \bigr] \bigl( 
3[{\bf n}.\hat{{\bf r}}]^2 - 1 \bigr) \tfrac{\hat{{\bf r}}}{r^2} \\
 + \bigl[ \tfrac{17a\xi^B\xi^F}{32} \bigr] \Bigl( \bigl( 1 - 3[{\bf n}.\hat{{\bf r}}]^2 \bigr) \tfrac{{\bf n}}{r^3} 
+ \bigl( 5[{\bf n}.\hat{{\bf r}}]^2 -3 \bigr) \tfrac{3[{\bf n}.\hat{{\bf r}}]\hat{{\bf r}}}{r^3} \Bigr) \Bigr\} \, 
,
\label{eq:flowfield}
\end{multline}
where $T$ is the period of the swimming stroke. For $\xi^F = \xi^B$ the swimmer is TP invariant and the dipolar 
term vanishes. (The amplitudes associated with each of the terms are given here only to lowest order in $\xi^B$, $\xi^F$. However 
all terms in the coefficient of $r^{-2}$ are antisymmetric in $\xi^F$, $\xi^B$ and hence vanish in the TP 
invariant limit.) 

The reason TP invariant swimmers do not have any dipole contribution can be seen as follows.   
If a general swimming motion is invariant under simultaneous time reversal and parity transformations then the 
average flow field must also have this property. The dipole term,
however, changes sign under a PT transformation and so the coefficient 
multiplying it must be zero. (This can be verified for the first term in Eq.~(\ref{eq:flowfield}).)

When the front and back arms have different amplitudes the TP invariance is broken and
both $r^{-2}$ and $r^{-3}$ contributions appear in the averaged far field flow in Eq.~(\ref{eq:flowfield}).
 A comparison of the magnitude of the two contributions shows that the dipolar term only 
dominates for distances 
\begin{equation}
\tfrac{r}{D} \gtrsim \tfrac{68}{29} \tfrac{D}{|\xi^B-\xi^F|} \, ,
\label{eq:flowscaling}
\end{equation}
which diverges when the amplitudes of the two arms become equal, as
required by the TP invariance. 

\begin{figure}
\begin{center}
\includegraphics[width=3.6in,height = 5.75in]{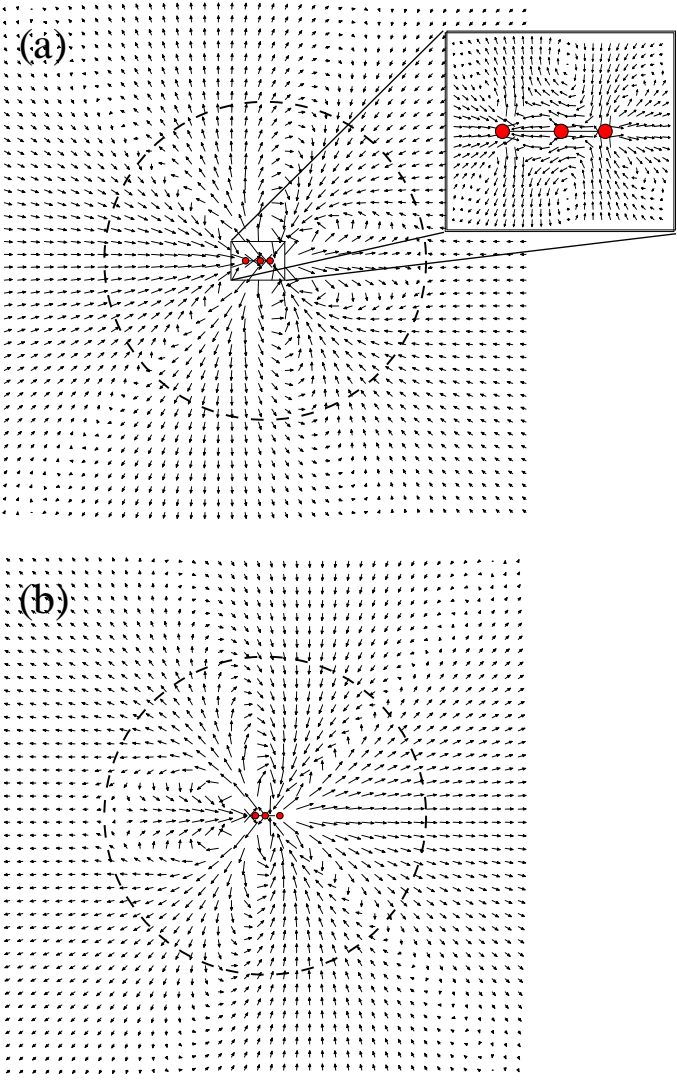}
\end{center}
\caption{The flow field around a swimmer averaged over one swimming cycle. (a) $\xi^B < \xi^F$ with the inset 
showing the detail near the swimmer, and (b) $\xi^B > \xi^F$. In both cases the direction of swimming is from left 
to right and the flow is axisymmetric about the swimmer. The magnitude of velocities are plotted on a logarithmic 
scale to enable both the near sphere and far field behaviour to be observed simultaneously. The parameters used 
were $a = 0.1D$ and $\xi^B$, $\xi^F = 0.2D$, $0.4D$.} 
\label{fig:velocity_field}
\end{figure}

The velocity field generated by this swimmer, averaged over a swimming cycle, is shown in 
Fig.~\ref{fig:velocity_field}. It was obtained using a numerical integration of the Oseen tensor equation, subject 
to the constraint that the swimmer is force and torque-free. Details of this method are given in \cite{david07}. 
Since the numerical technique does not exploit a far field expansion of the stokeslets it enables the flow near 
the swimmer to be determined as well, as illustrated in the inset. Close to the middle sphere the fluid is clearly 
seen to move to the left; essentially the swimmer acts like a pump which moves fluid from in front of it to behind 
it, leading to the swimmer moving to the right. The analytic far field expansion Eq.~\eqref{eq:flowfield} 
accurately describes (to within 10\%) the numerical flow for distances further than around three swimmer lengths. 
The dashed circle demarcates the estimated distance, Eq.~\eqref{eq:flowscaling}, at which the far field flow 
crosses over from $r^{-3}$ to $r^{-2}$ dipolar scaling.

It is interesting that two separate cases arise for the cross-over to a dipolar far field flow. This is because 
the $r^{-2}$ term is antisymmetric under $\xi^B \leftrightarrow \xi^F$ while the $r^{-3}$ term is symmetric, 
giving distinct flow fields for the two cases $\xi^B < \xi^F$ (Fig.~\ref{fig:velocity_field}(a)) and $\xi^B > 
\xi^F$ (Fig.~\ref{fig:velocity_field}(b)). As we shall show, this distinction results in significantly different 
long time behaviour of interacting swimmers. 

We now consider how the trajectory of a second nearby swimmer (Alice) will be altered by this flow field. The interaction 
will involve Alice both being advected by, and rotating in, the flow field generated by Bob. 
We find that there are two contributions to the interaction. The first, which we refer to as the {\em 
dead} term, is independent of the motion of Alice. This contribution may be calculated using the 
average flow field for a single swimmer, Eq.~\eqref{eq:flowfield}, which to leading order gives
\begin{equation}
\boldsymbol{\Delta x}_A^{\text{dead}} = \bigl[ \tfrac{29a^2\xi^B\xi^F(\xi^B-\xi^F)}{64D^2} \bigr] \Bigl( \tfrac{\hat{{\bf 
r}}}{r^2} \Bigr) \Bigl\{ 3[{\bf n}_B.\hat{{\bf r}}]^2 - 1 \Bigr\} \, , 
\label{eq:translatedead}
\end{equation}
\begin{multline}
\boldsymbol{\Delta \theta}_A^{\text{dead}} = \bigl[ \tfrac{87a^2\xi^B\xi^F(\xi^B-\xi^F)}{64D^2} \bigr] \Bigl( \tfrac{ 
 \hat{{\bf r}} \times {\bf n}_A}{r^3} \Bigr) \\
\times \Bigl\{ [{\bf n}_A.\hat{{\bf r}}] + 2[{\bf n}_A.{\bf n}_B][{\bf n}_B.\hat{{\bf r}}] - 5[{\bf n}_A.\hat{{\bf 
r}}][{\bf n}_B.\hat{{\bf r}}]^2 \Bigr\} \, ,
\label{eq:rotatedead}
\end{multline}
where subscripts $A,B$ label Alice and Bob, and ${\bf r}$ is the 
position vector of Alice relative to Bob. $\boldsymbol{\Delta x}_A$ represents the translation vector, and $\boldsymbol{\Delta \theta}_A$ the 
rotation vector (whose magnitude and direction define the rotation angle and axis respectively), induced in Alice as a result of the hydrodynamic interaction with Bob. Although the rotation is higher 
order in $r^{-1}$ it should not be considered less important. This is because small changes in angle can give rise 
to large positional displacements in the long time limit.

\begin{figure}
\begin{center}
\includegraphics[width=85mm,height=60mm]{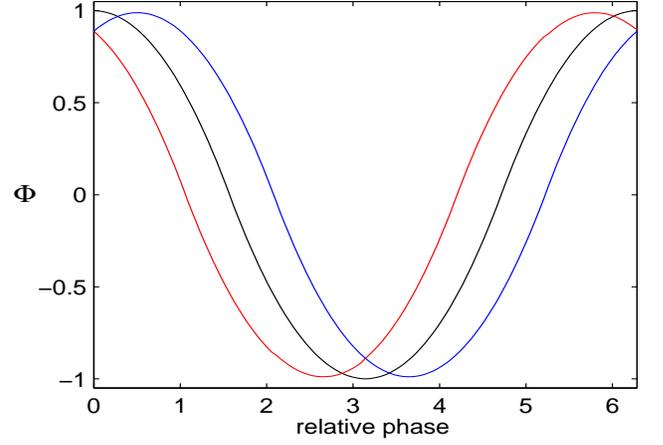}
\end{center}
\caption{The function $\Phi$ describing the effect of the relative phase of the swimmers on the interactions. The 
parameters used were: black line $\xi^B = \xi^F = 0.3D$, red line, $\xi^B = 0.4D, \xi^F = 0.2D$, and blue line, 
$\xi^B = 0.2D, \xi^F = 0.4D$.}
\label{fig:relphase}
\end{figure}

In addition, there is a {\em live} term which describes the more interesting interaction arising as a consequence 
of both organisms trying to swim. This term encodes all of the information about how the interactions depend on 
the relative phase of the swimmers. Using the Oseen tensor to describe the hydrodynamics we have determined the 
leading order contribution to the live interaction, finding a net swimming stroke translation and rotation given 
by
\begin{multline}
\boldsymbol{\Delta x}_A^{\text{live}} = \bigl[ \tfrac{3aD\xi^B\xi^F}{2}\Phi \bigr] \Bigl( \tfrac{\hat{{\bf r}}-[{\bf 
n}_A.\hat{{\bf r}}]{\bf n}_A}{r^3} \Bigr) \Bigl\{ [{\bf n}_A.\hat{{\bf r}}] \\
+ 2[{\bf n}_A.{\bf n}_B][{\bf n}_B.\hat{{\bf r}}] - 5[{\bf n}_A.\hat{{\bf r}}][{\bf n}_B.\hat{{\bf r}}]^2 \Bigr\}  
\, ,
\label{eq:translatelive}
\end{multline}
\begin{multline}
\boldsymbol{\Delta \theta}_A^{\text{live}} = \bigl[ \tfrac{-9aD\xi^B\xi^F}{8}\Phi \bigr] \Bigl( \tfrac{{\bf n}_A \times 
\hat{{\bf r}}}{r^4} \Bigr) \Bigl\{ 1 + 2[{\bf n}_A.{\bf n}_B]^2 - 5[{\bf n}_A.\hat{{\bf r}}]^2 \\
- 5[{\bf n}_B.\hat{{\bf r}}]^2 - 20[{\bf n}_A.{\bf n}_B][{\bf n}_A.\hat{{\bf r}}][{\bf n}_B.\hat{{\bf r}}] + 
35[{\bf n}_A.\hat{{\bf r}}]^2[{\bf n}_B.\hat{{\bf r}}]^2 \Bigr\}  \, .
\label{eq:rotatelive}
\end{multline}
The relative phase of the swimmers enters through the function $\Phi$. 
The precise form of $\Phi$ depends on the amount of time taken for each of the four steps in Fig.~\ref{fig:golepic}.
One choice is to assume that arm lengths vary at a constant rate $w$, so the time for steps (i) and (iii) is $\xi^B/w$, and for steps (ii) and (iv) is $\xi^F/w$. 
In this case $\Phi$ is plotted for a selection of 
parameter values in Fig.~\ref{fig:relphase}. Other choices, for instance requiring that each of the four steps takes 
equal time, produce very similar results.

When the two contributions to the interactions are compared we can see that the live contribution dominates for 
all distances
\begin{equation}
\tfrac{r}{D} \lesssim \tfrac{96}{29} \tfrac{D}{a} \tfrac{D}{|\xi^F-\xi^B|} \, \Phi \, .
\label{eq:interactionscaling}
\end{equation}
This behaves similarly to the average fluid flow in that it diverges as the swimmer becomes TP invariant. However, 
there is a crucial difference for non TP invariant swimmers. The additional factor of $D/a$ is large, and 
therefore ensures that for all distances over which the interactions are strong enough to be significant, the 
dominant contribution is from the live term, despite the fact that it is higher order in $r^{-1}$. Thus, in 
general, even though the average flow field may be dipolar, the interactions are not.

\begin{figure}
\begin{center}
\includegraphics[width=3in,height = 8in]{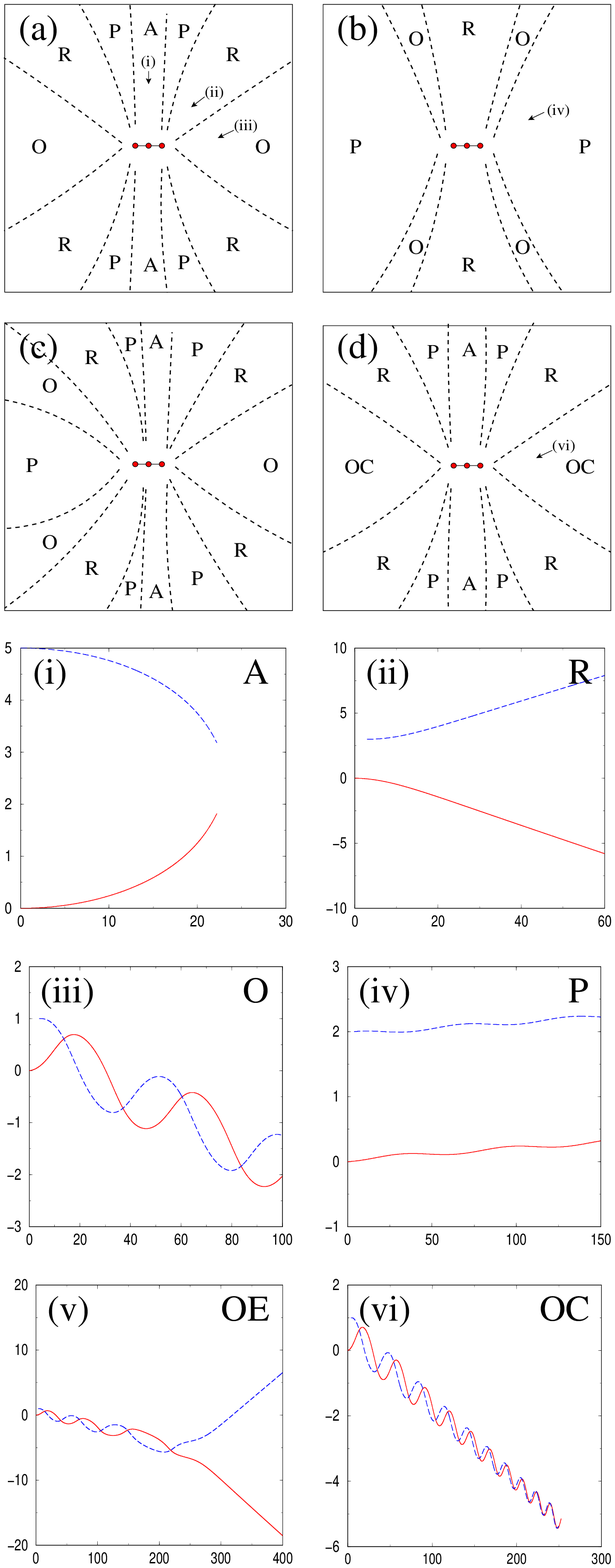}
\end{center}
\caption{Long time behaviour of two swimmers, Alice and Bob, (a) in phase, (b) $\pi$ out of phase, (c) with Alice phase lagging by $\pi/2$, and 
(d) in phase but with $\xi^F > \xi^B$. (i)--(vi) illustrate a selection of trajectories representing attractive 
(A), repulsive (R), oscillatory (O), parallel (P), expanding oscillatory (OE), and contracting oscillatory (OC) 
behaviour, respectively.}
\label{fig:angle}
\end{figure}

We turn now to an investigation of the long time behaviour resulting from these interactions by considering how 
the trajectories of two identical swimmers depend on their relative position, orientation and phase. We focus only 
on the most interesting case when the swimmers follow almost parallel trajectories since this allows time for the 
weak $r^{-3}$ interactions to have an appreciable effect. We consider first symmetric swimmers ($\xi^B=\xi^F$) 
swimming in phase, before looking at how the behaviour changes when these constraints are relaxed. The results are 
summarised in Fig.~\ref{fig:angle}(a). Alice initially lies at the origin (as shown in the figure) and Bob
is aligned parallel and displaced to the corresponding point in the diagram. The figure is divided into 
different regions, labelled A, R, O, and P, corresponding to the four different types of long term behaviour that 
were observed; attractive, repulsive, oscillatory and parallel trajectories.

Region A represents attractive behaviour, an example of which is illustrated in Fig.~\ref{fig:angle}(i). 
Initially, Alice lies at the origin and Bob is displaced $5D$ above her. The two curves show the centre 
of mass motion of the two swimmers in time. For accuracy, the trajectories were obtained numerically 
\cite{david07}, although the results which follow from iterating the analytic formulae, 
Eqs.~\eqref{eq:translatedead}-\eqref{eq:rotatelive}, are in close agreement. All scales are in units of $D$, the 
swimmer arm length. The parameters chosen were $a = 0.1D$ and $\xi = 0.3D$, giving a displacement per cycle of 
$\Delta \approx (7/12)a\xi^2/D^2 = 0.005D$. Thus, since the swimmers move a distance $\sim \! 25D$ before 
colliding, this simulation lasted for $\sim \! 5000$ swimming cycles, illustrating that the swimming trajectories 
evolve on much longer times scales than that of the individual swimming stroke. The lines terminate when the 
swimmers get too close, as here the Oseen tensor approximation breaks down. 

Region R contains swimmers which repel each other, as shown in Fig.~\ref{fig:angle}(ii). The swimmers each rotate 
through a small angle of $\sim \! 7^o$, with the rotation largely complete after $\sim \! 3000$ swimming cycles. 

Fig.~\ref{fig:angle}(iii) illustrates the oscillatory trajectories observed in region O. These oscillations take 
place over the course of $\sim \! 10000$ swimming cycles, although this period varies strongly with the 
separation of the two swimmers. It is notable that the trajectories exhibit a slow drift superimposed onto the 
oscillations.
The oscillations originate in the rotational part of the live interaction, while the superimposed drift is a 
consequence of the translational part. This highlights the significance of rotational interactions, despite their 
being higher order in $r^{-1}$.

Fig.~\ref{fig:angle}(b) shows the diagram corresponding to Fig.~\ref{fig:angle}(a) but with the two swimmers  
$\pi$ out of phase. The trajectories change because the amplitude of the live interaction changes sign. The 
attractive region is replaced by a repulsive one. The oscillatory regime now produces oscillating parallel paths, 
an example of which is shown in Fig.~\ref{fig:angle}(iv). Here the two swimmers follow parallel paths inclined at 
a small angle to the direction in which they would move in the absence of interactions. Both trajectories exhibit 
small amplitude, long period oscillations about the swimming direction; the oscillations of the two swimmers are 
exactly $\pi$ out of phase, and the amplitude of the oscillations vanishes when the angle between them 
is $\sim \! 30^o$, corresponding to one of the zeroes of $\boldsymbol{\Delta \theta}_A^{\text{live}}$.

Fig.~\ref{fig:angle}(c) shows the intermediate case when Alice has a $\pi/2$ phase lag with respect to Bob. 
This case is noteworthy as the amplitude for the lowest order live part of the interaction goes to zero. 
Nonetheless, it is clear that the live terms still dominate since the behaviour is different for swimmers placed 
in front or behind the central swimmer. This positional change is equivalent to changing the relative phase 
between the swimmers from $\pi/2$ to $3\pi/2$ and thus can only be important if the live terms are dominating the 
interaction.

The oscillations of Fig.~\ref{fig:angle}(iii) are perfectly periodic and repeat indefinitely. This is a direct 
consequence of the TP invariance of the swimmers when the amplitudes of the two arms are equal. When this invariance is 
broken, the oscillating states are no longer indefinitely stable. In particular, Fig.~\ref{fig:angle}(v) shows 
the effect of making the front swimming amplitude, $\xi^F$, smaller and the back amplitude, $\xi^B$, larger. Now 
the amplitude of the oscillations becomes successively larger until they can no longer be sustained and, finally, 
the two swimmers follow diverging trajectories. The opposite case,
when $\xi^F > \xi^B$, is shown in Fig.~\ref{fig:angle}(vi). The
amplitude of the oscillations decays and the swimmers move closer together, with the lines terminating when the 
Oseen tensor approximation breaks down as a result of the swimmers getting too close.
This behaviour is of particular interest as it suggests that suspensions of swimmers driven 
in phase might spontaneously start to form chains. 

In this article, we have considered the collective dynamics arising from hydrodynamic interactions 
between two model microswimmers. These interactions are not dominated by the leading non-zero term in a far field 
expansion of stokeslets, but rather by a higher order {\em live} term describing the simultaneous motion of both 
swimmers while they swim. 
Consequently, the long time behaviour is sensitive to the relative phase of the two swimmers and may be tuned by 
making small alterations to the details of the swimming stroke. 

Current experimental artificial microswimmers rely on external actuation in order to move~\cite{dreyfus05}. 
Therefore, situations in which two interacting swimmers are kept strictly in phase could be testable 
against the trajectories found in Fig.~\ref{fig:angle}.  
In biological systems, the individual swimmers move autonomously and the relative phases between 
them would be expected to change in time. Given the sensitivity of hydrodynamic interactions to 
relative phase, such changes, perhaps leading to phase-locking, may have a significant impact on collective 
swimmer motion and understanding this effect will be an important next step towards biological realism.

The behaviour of two swimmers provides only a flavour of the collective dynamics of larger groups of microorganisms.
Thus, although our analysis has shown that attractive hydrodynamic interactions exist between two swimmers, this fact alone 
is insufficient to claim clustering or swarming in larger groups of microorganisms. Similarly, the pairwise repulsion 
of two swimmers does not imply that larger groups will disperse. As such, there is much that still needs to be done 
to determine the collective behaviour of a large number of hydrodynamically interacting swimmers. This study 
has provided a framework in which this might be done and hinted at the richness which is likely to be found.

\begin{acknowledgements}
We thank M.E. Cates and R.E. Goldstein for helpful discussions and acknowledge
support from the EPRSC and ONR.
\end{acknowledgements}


\begin{thebibliography}{99}
\bibitem{purcell77} Purcell, E. M. Life at low Reynolds number. {\it Am. J. Phys.} {\bf 45}, 3-11 (1977).

\bibitem{lighthill76} Lighthill, J. Flagellar Hydrodynamics: The John von Neumann Lecture, 1975. 
{\it SIAM Review}, {\bf 18} (2), 161-230 (1976).

\bibitem{guirao07} Guirao, B. \& Joanny, J.F. Spontaneous Creation of Macroscopic Flow and Metachronal Waves in an Array of Cilia. 
{\it Biophys. J} {\bf 92}, 1900-1917 (2007). 

\bibitem{dreyfus05} Dreyfus, R. {\it et al.} Microscopic artificial swimmers. {\it Nature} {\bf 437}, 862-865 (2005).

\bibitem{higdon79b} Higdon, J.J.L. The hydrodynamics of flagellar propulsion: helical waves. 
{\it J. Fluid Mech.} {\bf 94}, 331-351 (1979).

\bibitem{shapere89a} Shapere, A. \& Wilczek, F. Geometry of self-propulsion at low Reynolds number. 
{\it J. Fluid Mech.} {\bf 198}, 557-585 (1989).

\bibitem{stone96} Stone, H.A. \& Samuel, A.D.T. Propulsion of Microorganisms by Surface Distortions. 
{\it Phys. Rev. Lett.} {\bf 77}, 4102 (1996).

\bibitem{shapere89b} Shapere, A. \& Wilczek, F. Efficiencies of self-propulsion at low Reynolds number. 
{\it J. Fluid Mech.} {\bf 198}, 587-599 (1989).

\bibitem{becker03} Becker, L.E., Koehler, S.A. \& Stone, H.A. On self-propulsion of micro-machines at low 
Reynolds number: Purcell's three-link swimmer. {\it J. Fluid Mech.} {\bf 490}, 15 (2003).

\bibitem{tam07} Tam, D. \& Hosoi, A.E. Optimal Stroke Patterns for Purcell's Three-Link Swimmer. 
{\it Phys. Rev. Lett.} {\bf 98}, 068105 (2007).

\bibitem{toner05} Toner, J., Tu, Y. \& Ramaswamy, S. Hydrodynamics and phases of flocks. 
{\it Ann. Phys.} {\bf 318}, 170-244 (2005).

\bibitem{goldstein04} Dombrowski, C., Cisneros, L., Chatkaew, S., Goldstein, R.E. \& Kessler, J.O.  
 Self-Concentration and Large-Scale Coherence in Bacterial Dynamics. {\it Phys. Rev. Lett.} {\bf 93}, 098103 (2004).

\bibitem{guell88} Guell, D.C., Brenner, H., Frankell, R.B. \& Hartman, H. Hydrodynamic forces and band formation 
in swimming magnetotactic bacteria. {\it J. Theor. Biol.} {\bf 135}, 525 (1988).

\bibitem{budrene91} Budrene, E.O. \& Berg, H.C. Complex patterns formed by motile cells of {\it Escherichia coli}. 
{\it Nature} {\bf 349}, 630 (1991).

\bibitem{budrene95} Budrene, E.O. \& Berg, H.C. Dynamics of formation of symmetrical patterns by chemotactic bacteria. 
{\it Nature} {\bf 376}, 49 (1995).

\bibitem{ramaswamy06} Ramaswamy, S. \& Simha, R.A. The mechanics of active matter: Broken-symmetry hydrodynamics 
of motile particles and granular layers. {\it Solid State Commun.} {\bf 139}, 617-622 (2006).

\bibitem{ishikawa06} Ishikawa, T. \& Hota, M. Interaction of two swimming {\it Paramecia}. 
{\it J. Exp. Biol.} {\bf 209}, 4452-4463 (2006).

\bibitem{kim04} Kim, M. \& Powers, T.R. Hydrodynamic interactions between rotating helices. 
{\it Phys. Rev. E} {\bf 69}, 061910 (2004).

\bibitem{nasseri97} Nasseri, S. \& Phan-Thien, N. Hydrodynamic interaction between two nearby swimming 
micromachines. {\it J. Comp. Mech.} {\bf 20}, 551-559 (1997).

\bibitem{pedley06} Ishikawa, T., Simmonds, M.P. \& Pedley, T.J. Hydrodynamic interaction of two swimming 
model micro-organisms. {\it J. Fluid Mech.} {\bf 568}, 119-160 (2006).


\bibitem{najafi04} Najafi, A. \& Golestanian, R. Simple swimmer at low Reynolds number: three linked spheres. 
{\it Phys. Rev. E} {\bf 69}, 062901 (2004).

\bibitem{pedley92} Pedley, T.J. \& Kessler, J.O. Hydrodynamic phenomena in suspensions of swimming 
microorganisms. {\it Ann. Rev. Fluid Mech.} {\bf 24}, 313-58 (1992).

\bibitem{ortiz05} Hernandez-Ortiz, J.P., Stoltz, C.G. \& Graham, M.D. Transport and Collective Dynamics in 
Suspensions of Confined Swimming Particles. {\it Phys. Rev. Lett.} {\bf 95}, 204501 (2005).

\bibitem{simha02} Simha, R.A. \& Ramaswamy, S. Hydrodynamic Fluctuations and Instabilities in Ordered 
Suspensions of Self-Propelled Particles. {\it Phys. Rev. Lett.} {\bf 89}, 058101 (2002).

\bibitem{david07} Earl, D.J., Pooley, C.M., Ryder, J.F., Bredburg, I. \& Yeomans, J.M. Modelling microscopic 
swimmers at low Reynolds number. {\it J. Chem. Phys.} {\bf 126}, 064703 (2007).
  
\end{thebibliography}
\end{document}